\documentclass[10pt, twocolumn]{article}

\usepackage[utf8]{inputenc}
\usepackage[left=0.55in, right=0.55in, top=0.6in, bottom=0.8in]{geometry}
\usepackage{graphicx}
\usepackage{amsmath, amssymb}
\usepackage{microtype}
\usepackage{titlesec}
\usepackage{authblk}
\usepackage{threeparttable}
\usepackage{wrapfig2} 
\usepackage{cuted}

\usepackage[authoryear, round]{natbib} 
\usepackage[colorlinks=true, allcolors=blue]{hyperref} 

\newcommand{\apj}{ApJ}
\newcommand{\apjl}{ApJL}
\newcommand{\apjs}{ApJS}
\newcommand{\mnras}{MNRAS}
\newcommand{\aap}{A\&A}

\usepackage{xspace}
\usepackage{amsmath}
\usepackage{gensymb}
\usepackage{ragged2e}

\newcommand{\fesc}{\mbox{$f_{esc}$}\xspace}
\newcommand{\lya}{\mbox{Ly$\alpha$}\xspace}
\newcommand{\sfrsd}{\mbox{$\Sigma_{\mathrm{SFRD}}$}\xspace}

\titleformat{\section}{\large\bfseries}{\thesection.}{0.5em}{}
\titleformat{\subsection}{\normalsize\bfseries}{\thesubsection.}{0.5em}{}
\titlespacing*{\section}{0pt}{1.5ex plus 1ex minus .2ex}{1ex plus .2ex}

\begin{document}

% --- 4. Manual Title Block (Full Width) ---
\twocolumn[
  \begin{@twocolumnfalse}
    \begin{center}
      {\LARGE \textbf{HST's Deep Blue: extremely deep UV imaging to reveal the contributors to reionization}} \par
      \vspace{1.5em}
      {\small Ilias Goovaerts$^1$, Alexander Beckett$^2$, Matthew Hayes$^3$, Alberto Saldana-Lopez$^3$, Annalisa Citro$^4$, Hakim Atek$^5$, Harry Teplitz$^6$, Roser Pello$^2$, Peter Watson$^7$, Claudia Scarlata$^4$, Mitchell Revalski$^1$, Davide Tornotti$^7$, Nicolas Laporte$^2$, Meriam Ezziati$^5$, Callum Witten$^8$} \par
      
      \vspace{0.5em}
      
      {\footnotesize $^1$\textit{Space Telescope Science Institute, 3700 San Martin Drive, Baltimore, MD 21218, USA}\\ 
      $^2$\textit{Aix Marseille Université, CNRS, CNES, LAM (Laboratoire d’Astrophysique de Marseille), 13388 Marseille, France}\\
      $^3$\textit{Stockholm University, Department of Astronomy and Oskar Klein Centre for Cosmoparticle Physics, AlbaNova University Centre, SE-10691, Stockholm, Sweden}\\
      $^4$\textit{Minnesota Institute for Astrophysics, School of Physics and Astronomy, University of Minnesota, 116 Church Street SE, Minneapolis, MN 55455, USA}\\
      $^5$\textit{Institut d'Astrophysique de Paris, CNRS, Sorbonne Universit\'e, 98bis Boulevard Arago, 75014, Paris, France}\\
      $^6$\textit{IPAC, California Institute of Technology, 1200 E. California Blvd, Pasadena, CA 91125, USA}\\
      $^7$\textit{Dipartimento di Fisica ``G. Occhialini'', Universit\'a degli Studi di Milano-Bicocca, Piazza della Scienza 3, 20126 Milano, Italy}\\
      $^8$\textit{Department of Astronomy, University of Geneva, Chemin Pegasi 51, 1290 Versoix, Switzerland}\par}
      \vspace{1.5em}
      
      \begin{minipage}{0.9\textwidth}
        \small
        Understanding galaxy evolution in the epoch of reionization and the effect these galaxies had on the transformation of the intergalactic medium from neutral to ionized, is a key goal of modern astrophsyics, and is central to both HST's and JWST's missions. The biggest remaining uncertainty is the escape fraction of ionizing photons from galaxies. Quantifying and understanding this at redshifts close to reionization is an objective only achievable through observations with HST, in particular,  deep imaging with the WFC3/UVIS instrument to detect ionizing photons from galaxies at $2<z<4$. A survey across 20 fields with supporting spectroscopy would both build up a sample similar to the state of the art at low redshift, and also overcome the uncertainty stemming from the unknown transmission of ionizing photons through the intergalactic medium. Such a program would establish the tracers of ionizing photon escape to use within the epoch of reionization and reconcile the growth of the first galaxies with the progression and topology of reionization. 
      \end{minipage}
      \vspace{1em}
    \end{center}
  \end{@twocolumnfalse}
]

\section{The Epoch of Reionization}\label{sect:context}

Understanding the epoch of reionization (EoR) has been a key goal of the \textit{Hubble} Space Telescope (HST) for the past decades, and continues to be a central goal of astrophysics in the current era of the \textit{James Webb} Space Telescope (JWST). A significant open debate remains: that of the contribution of ionizing sources to the process of reionization. The core difficulty in answering this question centres around the fraction of ionizing (Lyman continuum, LyC) photons, that escape from galaxies, \fesc, as these photons cannot be observed during the EoR due to absorption by neutral hydrogen in the InterGalactic Medium (IGM) along the line of sight.\\
\indent Ionizing emission must therefore be understood at lower redshifts ($z<4.5$) and indirect tracers established which can be used to quantify the ionizing properties of galaxies within the EoR. In particular, understanding ionizing escape at times close to the EoR is crucial, as while local Universe studies (up to $z\sim0.5$) benefit from the ability to conduct high-resolution spectroscopic studies, probe very low \fesc values \citep{Flury2022LyC_lowz_survey,Jaskot2024multivariateLyC_lowz} and even study spatially resolved escape  \citep{Komarova2024Haro11_spat_resolve,Ejdetjarn2026Haro11LyC_lya}, they are observed $\sim10\,\mathrm{Gyr}$ after the EoR. During this time many aspects of galaxy evolution have changed, for example star formation efficiency \citep{Madau1996SF_history,Harikane2025UVLF_spec}, merger rate \citep{Rodriguez-Gomez2015Illustris_mergers}, UV background \citep{Fan2006UVBackground_EoR}, galaxy metallicity \citep{Langeroodi2023Z_evolution} and interstellar medium (ISM) gas densities \citep{Topping2025e_densities_z_evolution}. Therefore, to answer the question of which population was responsible for the reionization process, these experiments must be extended to the highest redshifts possible, within hundreds of Myr of the EoR. 

\subsection{Lyman continuum emission at high redshift with HST}\label{subsect:LyC}
\vspace{-2mm}
HST has been at the forefront of this effort for a decade, and continues to be the only observatory capable of the observations necessary to detect and characterize faint ionizing emission at $z>3$ (within $1\,\mathrm{Gyr}$ of the EoR), thanks to its combination of sensitivity and spatial resolution, crucial advantages of being space-based, and having a large primary mirror. Its UV and blue-optical sensitivity ($2000-5000\,\mathrm{\AA}$) is also essential, as these are the wavelengths which detect LyC photons coming from $2<z<4$ galaxies.\\
\indent Ground-based efforts suffer from a high fraction of low-redshift interlopers \citep{Siana2015LyC_interlopers,Vanzella2018Ion3}, due to lower spatial resolution, and cannot access UV wavelengths due to atmospheric absorption. \\
\indent Investment in recent years has led to significant progress in understanding LyC escape at low redshift (see \citealt{Jaskot2025LyCreview} for a review). Multivariate predictors of \fesc can now be used \citep{Jaskot2024multivariateLyC_lowz}, taking into account the most reliable tracers which benefit from statistically meaningful samples of Lyman continuum emitters (LCEs) at $0.1<z<0.5$. The most promising tracers, which will benefit most from validation at $z>3$ are \textbf{1)} Lyman-$\alpha$ (\lya)-based tracers, such as \lya equivalent width and escape, the separation of double-peaked \lya lines \citep{FLury2022LzLCS_results} and the fraction of light emitted in the \lya halo \citep{SaldanaLopez2025HF}. The exact relationships are debated and have not yet been verified at high redshift \citep{Kerutt2024LAE_LyC,Citro2025LAEs_noLyC}.
\textbf{2)} Star formation rate surface density (\sfrsd), which correlates well with \fesc due to stellar feedback from concentrated star formation creating holes in the ISM for ionizing photons to escape \citep{Vanzella2018Ion3, Naidu2020UVbright_high_fesc, Flury2022LyC_lowz_survey,Marques-Chaves2022highfesc_LCE}.
\textbf{3)} The UV slope, $\beta$, which has shown a strong anticorrelation with \fesc \citep{Chisholm2022LyC_beta}; galaxies with bluer slopes, indicating less dust content, also show higher \fesc \textbf{4)} rest frame optical line diagnostics such as the $\mathrm{[OIII]/[OII]}$ ratio \citep{Izotov2018LyC_OIII/OII,FLury2022LzLCS_results}, encoding the ionization of the ISM, available at high redshift with JWST, and with future 30+ meter class ground-based telescopes. These tracers not only show promising correlations with \fesc, but are accessible and observable at all redshifts, ensuring that, if validated, they can be used deep into the EoR.

\subsection{The Intergalactic Medium}\label{subsect:IGM}
At $z\gtrsim2$, remaining islands of neutral hydrogen in the IGM absorb and scatter LyC photons along the line of sight \citep{Inoue2014IGM}. The exact fraction of emitted LyC emission that is absorbed along any given line of sight is not known. Current studies model a distribution of simulated IGM sightlines (e.g. \citealt{Bassett2021IGM_sim}) and derive a correction from this distribution, with associated uncertainties \citep{Beckett2025PIE,Goovaerts2026MXDFz4.4}. \\
\indent The only way to truly overcome this limitation is by observing enough sightlines to reduce the uncertainty associated with the IGM transmission correction averaged over the whole sample. \cite{Scarlata2025} shows 20 independent fields is the necessary amount.

In this white paper, we outline the observations necessary with HST to achieve the science goal of understanding the first galaxies' contribution to reionization by observing enough LCEs at $z>2$, showing that HST is the only instrument that can achieve this. Sect.~\ref{sect:Obs} gives the observational details and requirements, including possible observing strategies and targets. Sect.~\ref{sect:goals} then gives the science targets of these observations and their impact on our understanding of the early Universe. Throughout, we use the AB magnitude system \citep{OkeGunn1983} and assume a Salpeter IMF \citep{Salpeter1955}.

\section{Observations with HST}\label{sect:Obs}

\begin{figure*}
    \centering
    \includegraphics[width=0.9\linewidth]{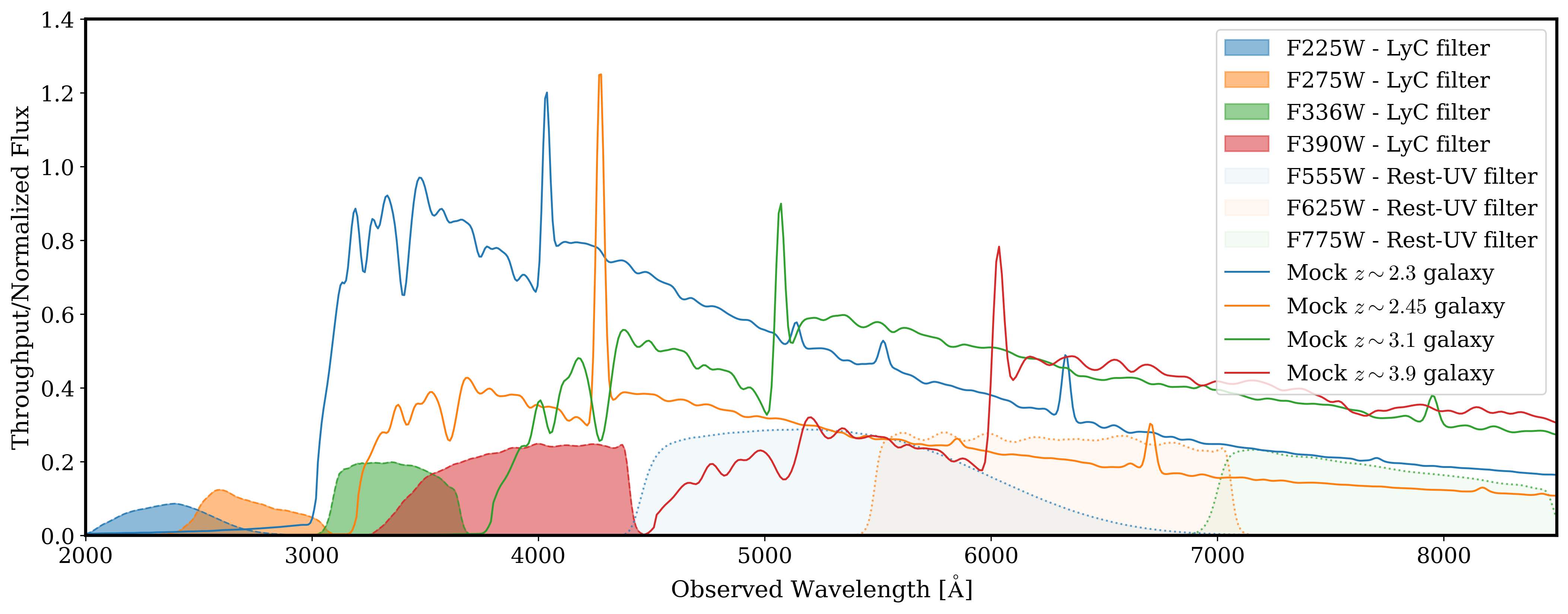}
    \caption{Four representative model spectra at the redshifts discussed in this work and the filters which probe their LyC emission. The filters which probe their rest frame UV emission are also shown. }
    \label{fig:3filters}
\end{figure*}

HST has already seen success in detecting and characterizing $z>2$ LCEs \citep{Vanzella2018Ion3, RiveraThorsen2022LyC_HUDF, Kerutt2024LAE_LyC, Mestric2025ion3,Goovaerts2026MXDFz4.4}, however, the vast majority of detections represent the extreme population of bright and very high-\fesc galaxies which are detectable with current observations. While these are valuable to study to gain insight as to the conditions which allow high \fesc, they are not representative of the galaxy population as a whole. An integrated understanding of \fesc down to low values is necessary to properly assess the galaxy contribution to reionization. The value often quoted as a population-average necessary for reionization is $5\,\%$ \citep{Finkelstein2019reionization_fesc}. This is therefore the ultimate goal at $z>3$, to be able to study these low levels of \fesc for a statistically meaningful sample of galaxies, spanning a range of stellar masses and SFRs. We devote the rest of this section to outlining precisely the future HST observations required to achieve this goal.  \\
\indent The aim is to build up a sample of a similar, ideally larger size than the current state of the art at low redshift (\textit{Low-z Lyman Continuum survey (LzLCS), $\sim50$ LCEs}; \citealt{Flury2022LyC_lowz_survey}), anticipating the improvement of this sample over coming years. In particular, as mentioned, a strong focus is required on the low-\fesc regime, as at low redshift, many LCEs are detected and characterized down to the percent level, making comparison to LCEs with many times more LyC escape than this of limited use. Therefore, \fesc levels down to 5\% must be targeted, for faint galaxies. \\
\indent We outline the possibilities for four different filters targeting LyC from galaxies at different redshift ranges, and compare them in Table~\ref{tab:filters}. F225W, F275W, F336W and F390W target LyC in the redshift ranges $z>2.3$, $z>2.45$, $z>3.09$ and $z>3.9$ respectively. Higher energy ionizing radiation is detected in each of these filters for higher redshifts than those quoted. This is also important to constrain the shape of the ionizing spectrum and further understanding of energetic extreme UV radiation.\\
\indent Each filter has specific advantages, however to present a one-to-one comparison, we consider the number of orbits necessary for each filter to detect 50\% intrinsic LyC escape for galaxies of $\mathrm{M_{UV}}\lesssim-19.5$ at $3\sigma$ significance, which, assuming standard dust-free conversions to SFR and using the galaxy main sequence from \citet{Popesso2023galaxy_MS} equates to galaxies of $>\mathrm{M_{*}\sim10^9\,M_{\odot}}$. Such an observing depth would also detect higher levels of escape from fainter galaxies (e.g., taking the case of F336W, $75\,\%$ escape from a $\mathrm{M_{UV}}\sim-19$ galaxy) and lower levels of escape from brighter galaxies (e.g. $10\,\%$ escape from a $\mathrm{M_{UV}}\sim-21$ galaxy). This mass range is just above the generally accepted ``dwarf-galaxy'' regime, and firmly into the ``faint'' category that previous studies have considered when assessing the galaxy contribution to reionization (with the ``bright'' category starting at $\mathrm{M_{UV}}\sim-23$) \citep{Robertson2015reionization_gals,Mascia2024EoR_reionizers}. We discuss how to access even fainter regimes in Sects.~\ref{subsubsect:lensing} and \ref{subsect:HWO}. This limit also corresponds well to the current limits of JWST within the EoR (see, e.g. \citealt{Mascia2024EoR_reionizers}). Lower intrinsic \fesc can be reached through stacking galaxies, for example 100 galaxies to reach the target $5\%$, easily achievable over 20 fields. \\
\indent Bluer filters probe LyC further from the EoR, hence have less relevance for this epoch, however they suffer from less IGM attenuation, allowing larger samples to be constructed. F336W and F390W probe LyC close to the EoR (within 1 Gyr), however, suffer from significant IGM attenuation along the majority of sightlines. On the other hand, these filters have high throughputs and steep red cutoffs, which make them attractive for the purpose of detecting faint LyC emission. Both have already been used for this purpose \citep{Kerutt2024LAE_LyC,Vanzella2018Ion3,Mestric2025ion3} and further efforts with F336W have been accepted as part of an effort to observe LyC from faint galaxies in the MUSE eXtremely Deep Field \citep{bacon2022musedatareleaseII}: \textit{HyperDeepUV}, (HST PID: 18004) and from brighter galaxies across 20 fields: \textit{High-z Lyman Continuum Survey (HzLCS)} (HST PID: 18080).\\
\textit{HyperDeepUV} will constrain \fesc from faint galaxies at $z>3$ (expecting to deliver constraints on \fesc for up to $\simeq 1500$ faint galaxies) and \textit{HzLCS} will overcome cosmic variance and uncertainty due to IGM absorption thanks to its 20 fields \citep{Scarlata2025}. We advocate for observations with HST that combine these two strategies, enabling the characterization of faint LCEs across 20 independent fields. \\
\indent Such an observing program would probe intrinsic \fesc down to $50\%$ for a sample of $~\sim30$ galaxies per UVIS field of view \citep{Reddy2009UVLF}: $\sim600$ across the 20 fields (also probing intrinsic \fesc down to $10\%$ for $\sim70$ galaxies). Constructing a sample similar to the \textit{LzLCS} in number will then be within reach, in addition to stacking analysis to reach very low \fesc values. We note that spectroscopy of all these galaxies is necessary to confirm their redshifts, and give routes to obtaining such spectroscopy or leveraging existing fields in the following section.

\begin{table*}[t] 
\centering
\begin{threeparttable} 
\caption{The four HST filters targeting Lyman continuum emission at different wavelengths, together with their particular advantages for use in an observing campaign.}
\vspace{1mm}
\label{tab:filters}
\begin{tabular}{|p{1.8cm}|p{1.7cm}|p{2cm}|p{1.9cm}|p{3.5cm}|p{4.5cm}|} 
\hline 
\textbf{LyC Filter} & \textbf{LyC exp. time}\tnote{a} & \textbf{Rest frame UV filter} & \textbf{LyC range}\tnote{b} & \textbf{IGM transmission}\tnote{c} & \textbf{Advantages} \\
\hline 
\hline
F225W & 90 orbits & F555W F625W\tnote{d} & $2.3$$<$$z$$\lesssim$2.8 & $\mathrm{mean = 9.9\%}$, 1$\sigma$ range 1.9\% - 12.2\%\tnote{e}    & Low effective IGM attenuation\tnote{e}.   \\
\hline
F275W & 50 orbits & F555W F625W\tnote{c} & $2.45$$<$$z$$\lesssim$$3.0$ & $\mathrm{mean = 16.1\%}$, 1$\sigma$ range 4.3\% - 20.8\%    & Lowest effective IGM attenuation: larger samples/less exposure time  \\
\hline 
F336W  & 130 orbits & F625W & $3.09$$<$$z\lesssim$$3.6$ & $\mathrm{mean = 11.8\%}$, 1$\sigma$ range 1.5\% - 14.8\%     &  Closer to EoR. Higher throughput and steep red cutoff. Low effective IGM attenuation. MUSE-detected \lya.   \\
\hline 
F390W & 1000 orbits & F775W & $3.9$$<$$z\lesssim$$4.3$  & $\mathrm{mean = 2.8\%}$, 1$\sigma$ range 0.3\% - 3.3\%     & Closest to EoR. High throughput and steep red cutoff. MUSE-detected \lya.   \\
\hline 
\end{tabular}

\begin{tablenotes}
   \item[a] Approximate exposure times needed to detect an intrinsic \fesc at the 50\% level for a galaxy of $\mathrm{M_{UV}=-19.6}$, taking into account mean IGM attenuation (see text). 
   \item[b] Upper range limits indicate the most efficient redshift range to detect LyC, it is possible to detect LyC above these depending on brightness, escape and shape of ionizing spectrum. Lower range limits are imposed by the LyC filter wavelength coverage.
   \item[c] Mean and $\pm1\sigma$ interval for IGM transmission, convolved with LyC filter; modelled using the TAOIST-MC code described in \cite{Bassett2021IGM_sim}.
   \lya redshifts into F555W at $z=2.6$ so F625W is preferable above this redshift.
   \item[d] \lya redshifts into F555W at $z=2.6$ so F625W is preferable above this redshift.
   \item [e] Filter shape means effective IGM transmission is lower, despite IGM opacity decreasing at $z=2.3$ versus $z=2.45$.
\end{tablenotes}
\vspace{-3mm}
\end{threeparttable} 
\end{table*}

\vspace{-3mm}
\subsection{The need for spectroscopy}\label{sect:spec}
\begin{figure}[h!]
    \centering
    \includegraphics[width=\linewidth]{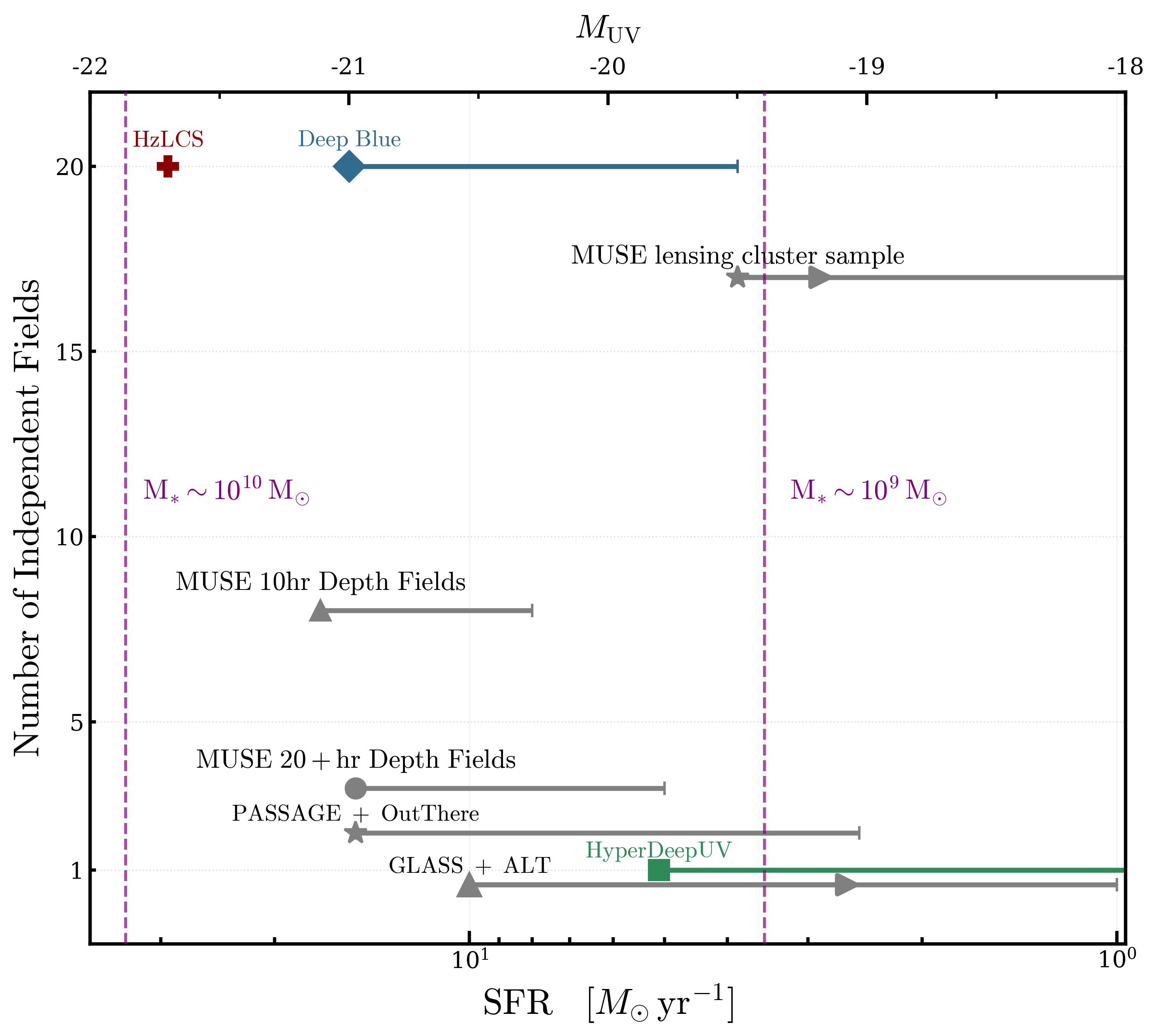}
    \vspace{-6mm}
    \caption{Currently available fields with deep spectroscopy for contribution to Deep Blue; the observational goal discussed in this work. Data points and error bars represent the mean and lower limit of accessible SFRs in each field based on spectroscopic depth. Grey points represent available fields without (sufficient) imaging targeting LyC. Recently accepted observing programs \textit{HyperDeepUV} and \textit{HzLCS} are coloured. Those with right-pointing arrows represent lensing fields where SFR depends on magnification. Gravitational lensing can extend this to $\mathrm{M_{UV}\sim-16}$ with moderate magnification. SFR-mass conversion from \cite{Popesso2023galaxy_MS}. Information on the various fields comes from \cite{Treu2022GLASS, Glazebrook2023cy2_proposal, Claeyssens2022LLAMAS, bacon2022musedatareleaseII, Naidu2024ALT, Tornotti2025MUDF_LAEs, Huberty2026PASSAGEspec_cat}.}
    \vspace{-2mm}
    \label{fig:fields}
\end{figure}

\vspace{-2mm}
Detecting LyC using HST imaging requires reliable galaxy redshifts, to make sure a LyC detection by HST is not due to a low-redshift interloper emitting non-ionizing flux in the filter. Therefore, the most economical strategy is to target fields with existing spectroscopy. Appropriate fields must have the required spectroscopic depth and wavelength coverage to detect emission lines from the faint $z>3$ galaxies which are targeted by the HST observations. \\
\indent The most efficient observations to use are IFU or Wide Field Slitless Spectroscopy (WFSS). The only IFU capable of fulfilling the aforementioned needs, with a large enough field of view, is the Multi-Unit Spectroscopic Explorer (MUSE) on the Very Large Telescope in Cerro Paranal, Chile. The only current WFSS instruments similarly capable are those aboard JWST: NIRCam and NIRISS's WFSS modes. Both these modes provide spectra for all the sources in their field of view, a crucial advantage which ensures a flux-limited sample of redshifts free from selection biases (other than selection by the emission lines used to confirm the redshifts).   \\
\indent MUSE detects \lya above $z=2.9$, so is only useful for LyC searches above this value. JWST/NIRISS's F200W filter detects $[\mathrm{OIII}]$ and H$\beta$ at this redshift range (e.g. appropriate for LyC searches with F336W) and detects it at relevant redshifts for searches with F225W and F275W in a combination of F150W and F200W (although the filter gap means redshifts are challenging to achieve with the $[\mathrm{OIII}]$ and H$\beta$ lines at $z\sim2.4$).\\
\indent There exists a rich archive of MUSE and WFSS data (see Fig.~\ref{fig:fields}) which is suitable for this purpose. Around 10 hours of exposure time with MUSE is sufficient for a robust, $>10\sigma$ detection of \lya emission from all the galaxies from which LyC will be detectable. A $>10\sigma$ detection is desirable for two reasons, firstly to confirm the line as \lya with its asymmetric shape, which becomes easier to identify in higher SNR lines. Secondly, \lya morphology has been shown to correlate well with \fesc \citep{SaldanaLopez2025HF,Goovaerts2026MXDFz4.4} and $>10\sigma$ is the minimum integrated SNR to enable morphological analyses. \\
\indent The deepest current WFSS observations have limiting line fluxes similar to those necessary to detect the LyC-emitting galaxies (see Fig.~\ref{fig:fields} and \citealt{Boyett2022GLASS_WFSS,Glazebrook2023cy2_proposal,Pirzkal2024NGDEEP,Naidu2024ALT,Huberty2026PASSAGEspec_cat}). Redshift confirmation from the $[\mathrm{OIII}]$ and H$\beta$ lines is preferred to H$\alpha$, as single line emitters can have significant redshift confusion. The H$\beta$ flux sets the limiting factor and gives an SFR estimate in this case. JWST WFSS observations, targeting these lines are necessary at $z<2.9$ as this is the MUSE redshift desert \citep{bacon2022musedatareleaseII}.

%\vspace{9mm}

\vspace{-3mm}
\subsection{Observational strategies for HST}\label{subsect:obs_strat}
\vspace{-2mm}

Observing 20 fields to the depth necessary depends on the redshift range and hence imaging filter chosen (Table~\ref{tab:filters}), starting from $\sim1000$ orbits with F275W. Additional time in the rest frame UV filter is necessary (see Table~\ref{tab:filters}) if it does not already exist. This is in order to detect the rest frame UV (non-ionizing) photons of the target galaxies ($\lambda\sim1500\,\mathrm{\AA}$), necessary to calculate the escape fraction ($f_{LyC}/f_{UV}$). This is far less expensive, achieved in $5-10$ orbits per field of the filters which see the rest frame UV flux for each redshift range. Note that these filters should avoid contamination from the \lya line. \\
\indent Given the large number of orbits needed for the requisite depth and number of fields, this is unachievable through the normal PI-led GO programs. Instead, a large-scale effort such as a 500-1000 orbit treasury or DDT program is recommended. This would also homogenise observations, facilitating analysis.\\
\indent This science case therefore necessitates continued full support for the WFC3/UVIS instrument. ACS is useful as observing the rest frame UV of these galaxies is cheaper with ACS due to higher throughputs. Detailed SED fitting will be necessary to fully constrain these galaxies' properties \citep{Kerutt2024LAE_LyC,Mestric2025ion3,Goovaerts2026MXDFz4.4}, so imaging with multiple HST filters spanning the full wavelength coverage (as well as other deep space-based imaging) is important for a full science exploitation.  

\subsubsection{Gravitational Lensing}\label{subsubsect:lensing}
\vspace{-2mm}
The magnification from strong gravitational lensing can be used to access even fainter regimes \citep{Soucail1988lensing_arc,kneib1996lenstooloriginal,jullo2007lenstool,Atek2025GLIMPSE}. MUSE has observed a number of lensing clusters \citep{richard2021MUSElensingclusters}, including some deep observations of clusters with large areas of magnification $>2$. A magnification of 2 halves the luminosity that would be detectable in the aforementioned exposure time, meaning the same analysis can be carried out for galaxies of $\mathrm{M_{UV}}\sim18.8$. Using regions of higher lensing magnification, e.g. $5+$, would allow the $\mathrm{M_{UV}}\sim -16$ to $-17$ regime to be reached \citep{Jung2024lensed_LCEs}, albeit in a smaller survey volume.\\
\indent Additionally, a huge number of new strong lensing clusters will be revealed in the coming years by surveys carried out by the Nancy Grace Roman Space Telescope and ESA's Euclid mission \citep{Atek2025Euclid_lensingclusters,Cerny2026SLICE}. \\
\indent Uncertainties from lens modelling can be well mitigated in well-studied clusters with many multiple images to constrain the lens models and by limiting analysis to regions with moderate magnification $<10$ \citep{Limousin2016lensing_uncertainties,richard2021MUSElensingclusters}.\\
\vspace{-2mm}

\section{Science Goals}\label{sect:goals}

The science goals that can be achieved with these deep UV observations are numerous. We list and describe the main goals, as pertain to the overarching goal of a better understanding of early galaxy evolution and the EoR. 

\vspace{-2mm}
\subsection{Reconciling the first galaxies with the reionization process}\label{subsect:firstgals}
\vspace{-1mm}
In recent years, ``photon budget crises'' have been discussed \citep{Munoz2024photonbudget,Simmonds2024bursty_gals_ionizing,Mascia2024EoR_reionizers}: scenarios in which there are too few or too many ionizing photons available from the first galaxies ($6<z<14$) for reionization follow the timeline established by \cite{Fan2006UVBackground_EoR,McGreer2015reionization_timeline,Planck2016reionization}. Early, late, patchy reionization scenarios have all been discussed, however these studies have been carried out without meaningful constraints on the escape fraction of these ionizing photons from the first galaxies into the IGM. By establishing a sample of LCEs at high redshift, reliable, statistically significant trends can be built up that can be safely extrapolated to earlier times, due to the short elapsed time between $z\sim3$ and $z\sim6$.\\
\indent Furthermore, with a sample of hundreds of LCEs, an ionizing luminosity function can be constructed and compared to the UV luminosity function \citep{Reddy2009UVLF,Bouwens2015ionizing_luminosity_density}. This will be a guide to all future LyC observations, reveal crucial insights into the relationship between the ionizing and non-ionizing continua and form a starting point to calculate the UV background, which is presently unconstrained.

\vspace{-2mm}
\subsection{ISM and CGM morphological impacts on \fesc}\label{subsect:ISMCGM}
\vspace{-1mm}

A crucial advantage of HST targeting fields with IFU observations from the MUSE instrument is access to spatial and spectral \lya properties for the LCEs. Both are promising \fesc tracers at low redshift, for example the \lya line separation \citep{FLury2022LzLCS_results} and \lya halo fraction \citep{SaldanaLopez2025HF}. The \lya emission line encodes key insights into the physical properties of the ISM and CGM of galaxies, and can be used to link this behaviour to the mechanisms for LyC escape. A simple, direct pathway for LyC photons out of the ISM and CGM of the host galaxy engenders high \fesc. This condition is recognizable by \lya emission emitted at the systemic redshift, indicating minimal scattering, and the majority over the core of the stellar, UV light (a low halo fraction). Understanding these mechanisms at high redshift will give the most promising path forward to quantifying \fesc within the EoR. 

\vspace{-2mm}
\subsection{Extreme, compact star-forming galaxies}\label{subsect:SFGs}
\vspace{-1mm}

Strong LCEs often exhibit compact, highly star forming regions \citep{Vanzella2018Ion3,Naidu2020UVbright_high_fesc,Marques-Chaves2022highfesc_LCE}. Studying these extreme galaxies will give insight as to the relationship between star formation and LyC production escape. This is likely of paramount importance due to the twofold impact of young, massive stars \citep{Goovaerts2026MXDFz4.4}. Ionizing photon production is heightened from the hottest, youngest, most massive stars \citep{Elridge2008BPASS,Hawcroft2025pySTARBURST}, and stellar feedback from these extreme populations clears the ISM for these ionizing photons to escape.

\vspace{-2mm}
\subsection{Preparing for the Habitable Worlds Observatory}\label{subsect:HWO}
\vspace{-1mm}
Achieving the above science goals, particularly \ref{subsect:firstgals}, will not only help JWST achieve its core science objective of understanding galaxy formation in the EoR, but will pave the way for the Habitable Worlds Observatory (HWO) to push similar analysis into the ultra faint and dwarf galaxy regime (down to ($\mathrm{M_{UV}\sim-13}$ -- see \citealp{Citro+2026hwo}). Not only will an established ionizing luminosity function guide all future observations of ionizing sources (guiding depth and area requirements), but the breadth of the program outlined herein will uncover the most interesting targets for specialised follow-up, for example with the UV IFU aboard HWO. Finally, an ionizing luminosity function, as well as all the physics probed with excellent statistics by the strategies recommended, would inform the next generation of Lyman continuum and reionization simulations, ensuring they make testable predictions for HWO. \\

\indent In summary, fulfilling the science goal of understanding galaxy evolution within the EoR and the galaxy connection to the process of reionization requires continued support for the WFC3/UVIS instrument, and would benefit from the ACS instrument and a large multi-cycle 500-1000 orbit treasury program. HST remains the only observatory capable of observing ionizing photons from faint galaxies close to the EoR and can make both revolutionary advances in the coming decade and pave the way for HWO in the 2040s and beyond. 

%\begin{itemize}
%    \item The need to reconcile the growth of first galaxies with the changes they made to the IGM. If there are too many ionizing photons, implications for clumping factor. If there aren't enough, are we missing galaxies? 
%    \item Related to above, the \textbf{timeline} of reionization. We still don't know when the midpoint was. 
%    \item ISM and CGM morphology and its impact on lyC fesc. ``Uncovering the journey of LyC photons through the ISM and CGM''.
%    \item Studying the most extreme, compact SFGs. 
%    \item what percentage of LAEs are LCEs? Simple measure to use within the EoR. (complicated by the IGM...)
%\end{itemize}

%We need more fields to overcome correlation of IGM and cosmic variance. There are many MUSE fields with 10h+ of data. Look what XX hours of F336W (and F435W) would accomplish. 

%Describe what kind of programs can achieve this. Big DDT? Either community or DDT. 

%Enable JWST to do its core science function. 
%What about HWO? How can HST support HWO. What will be most impactful science for HWO. What do we need from HST now to write the C1 HWO programs I want to write?

%Mention Par682 Glazebrook Cycle 2 - include Peter?

% --- Bibliography ---
\bibliographystyle{abbrvnat}

%\bibliographystyle{abbrvnat} 
%\bibliography{deepblue_new}

\end{document}